\documentclass[varenna,tikz]{cimento}
\usepackage[latin1]{inputenc}
\usepackage[T1]{fontenc}
\usepackage[english]{babel}
\usepackage{amssymb,amsmath} 
\usepackage{graphicx}
\usepackage{pgfplots} 
\usepackage[mode=buildnew]{standalone}
\usepackage{siunitx}
\usepackage{ulem}

\newcommand{\bs}{\boldsymbol}
 \newcommand{\D}{{\rm d}}
 \newcommand{\I}{{\rm i}}
 \newcommand{\E}{{\rm e}}
 \newcommand{\xibath}{\xi}
 \newcommand{\NT}{N_{\rm T}}
 
\title{Townes soliton and beyond: \\
Non-miscible Bose mixtures in 2D}

\author{Brice Bakkali-Hassani$^{1}$ and Jean Dalibard$^{2}$}

\institute{$^{1}$Department of Physics, Harvard University, Cambridge, Massachusetts 02138, USA}
\institute{$^{2}$Laboratoire Kastler Brossel,  Coll\`ege de France, CNRS, ENS-PSL University, Sorbonne Universit\'e, 11 Place Marcelin Berthelot, 75005 Paris, France}

\begin{document}

\maketitle

\begin{abstract}
In these lecture notes, we discuss the physics of a two-dimensional binary mixture of Bose gases at zero temperature, close to the point where the two fluids tend to demix. We are interested in the case where one of the two fluids (the bath) fills the whole space, while the other one (the minority component) contains a finite number of atoms. We discuss under which condition the minority component can form a stable, localized wave packet, which we relate to the celebrated "Townes soliton". We discuss the formation of this soliton and the transition towards a droplet regime that occurs when the number of atoms in the minority component is increased. Our  investigation is based on a macroscopic approach based on coupled Gross-Pitaevskii equations, and it is complemented by a microscopic analysis in terms of bath-mediated interactions between the particles of the minority component. 
\end{abstract}

\begin{figure}[t]
\begin{center}
\includegraphics{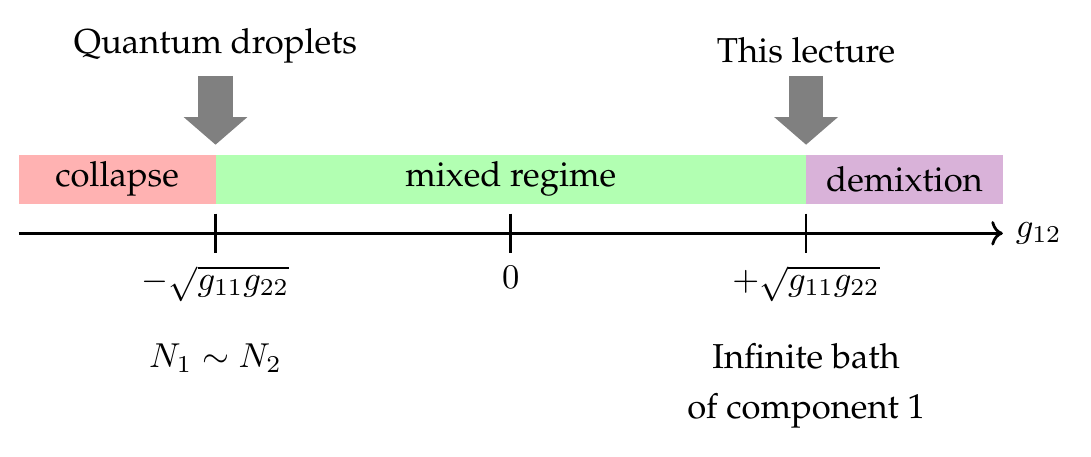}
\end{center}
\caption{Generic phase diagram for a binary mixture as function of the intercomponent interaction strength $g_{12}$, assuming repulsive intraspecies ($g_{ii}>0$). This lecture is devoted to the 2D case. The study of \cite{2015_petrov} regarding the generation of quantum droplets stabilized by beyond-mean-field forces addresses the 3D case.}
\label{fig:axis_g}
\end{figure}

Binary mixtures of low-temperature Bose gases can lead to a large variety of phenomena, depending on the strength and nature -- repulsive or attractive --  of intraspecies and interspecies interactions \cite{2016_pitaevskii,2008_pethick}. Even when each component is individually stable at the mean-field level, {\it i.e.}, it has a positive scattering length, a demixing instability, resp.\,a collapse, may occur if the interspecies interaction is sufficiently large and repulsive, resp.\, sufficiently large and attractive (see figure \ref{fig:axis_g}). 

The vicinity of these two singular points is particularly interesting. For example, the regime of  quantum droplets in three dimensions predicted in \cite{2015_petrov} occurs close to the collapse threshold. It takes advantage of the smallness of the mean-field interaction energy at this point: Beyond-mean-field corrections can play a significant role and they give access to a liquid state of matter, although the density remains several orders of magnitude lower than in usual liquids.

In this lecture, we focus on the vicinity of the demixing point. We consider the case where one component forms an infinite bath filling the entire space, while the second component contains only a finite number of particles $N_2$. By contrast to \cite{2015_petrov}, all the effects considered in these notes originates from mean-field interactions. In addition, we suppose that the system is at zero temperature. The state of the binary mixture is thus well described by two coupled Gross-Pitaevskii equations.

We wish to address the general question of the equilibrium shape of the minority component in these conditions: does it fill the whole space as the majority component, or does it form a stable, localized wave packet immersed in the bath? In the latter case, is it possible to describe this localized state using a (modified) Gross-Pitaevskii equation involving only the minority component?  

More specifically, we are interested here in the two-dimensional (2D) case.  For given interaction strengths, we show that there exists a threshold value $N_{\rm T}$ for the number of particles in the minority component, below which this component tends to fill the whole space. For any $N_2 > N_{\rm T}$, the minority component can form a stable wave packet with a definite size. When $N_2$ decreases to $N_{\rm T}^+$, the minority component converges towards a stationary wave packet that constitutes a realization of the celebrated "Townes soliton" \cite{1964_chiao}. This solitary wave is a remarkable example of a scale-invariant object, which can exist in principle with an arbitrary size. On the contrary, as $N_2$ increases towards values much larger than $N_{\rm T}$, the wave packet gradually evolves towards a droplet-like object with a density imposed by the bath.

The outline of this lecture is the following. In \S\,\ref{sec:solitons_2D}, we briefly review how solitons can be formed in a fluid described by the Gross-Pitaevski energy functional. We explain the special feature of the 2D case in relation with scale invariance. Then, we briefly describe two recent experiments in which Townes solitons were observed with a 2D cold atom setup \cite{chen2021observation,2021_bakkali}. In \S\,\ref{sec:binary_mixture}, we present a theoretical modeling of the binary mixture starting with the coupled Gross-Pitaevskii equations associated to each component. We discuss various situations in which the degrees of freedom of the bath can be eliminated to the benefit of a single  equation for the minority component. We explain how the system evolves when the number of particles is increased above the Townes threshold $N_{\rm T}$ and we describe the transition towards the droplet regime, where the minority component sits in a localized region of space where the bath is fully depleted. In \S\,\ref{sec:microscopic}, we turn to another point of view on this system and study the interactions between the particles of the minority component that are mediated by the bath. We first address the case where the mass  of a particle of the minority component $m_2$ is much larger than the mass of a bath particle $m_1$. We show that in this case, the mediated interactions are well described by a Yukawa potential, with a range related to the bath healing length. Then, for momenta $k$ of the minority component much smaller than $1/\xi$, we use Born approximation to simplify the description of these mediated interactions, and show that the result can be extended to the case $m_1=m_2$. This allows us to recover the results obtained in \S\,\ref{sec:binary_mixture} by the macroscopic approach. Finally, we draw  in \S\,\ref{sec:conclusions} some conclusions and perspectives regarding this two-dimensional binary mixture of Bose gases.


\section{Solitons in two dimensions}
\label{sec:solitons_2D}

\subsection{The Gross-Pitaevskii energy functional}

A soliton is an emblematic object of non-linear wave physics \cite{dauxois2006physics}. It is defined as a wave packet that maintains its shape over time, as a result of the competition between non-linear and dispersive effects. In this lecture, we consider a wave packet $\psi(\bs r,t)$ whose energy at time $t$ is described by the Gross-Pitaevskii functional in $D$ spatial dimensions:
\begin{equation}
E[\psi]\ =\frac{1}{2}\int \left( \left|\boldsymbol \nabla \psi(\bs r,t)\right|^2 \ +\  g \left|\psi(\bs r,t)\right|^4\right)\ {\rm d}^D r.
\label{eq:GP_functional}
\end{equation}
This energy functional is relevant in optics (for the propagation of laser beams in a non-linear medium), in atomic physics (for the classical-field description of a weakly-interacting Bose gas) and in condensed matter (e.g., as an order parameter for superfluid liquid helium). Note that we have set here $\hbar=1$ and $m=1$ for the mass of the particles, in the case where $\psi$ describes a matter-wave field. The first contribution to (\ref{eq:GP_functional}) corresponds to diffraction or kinetic energy, and the second one to the (cubic) non-linearity of the medium or to the interactions between  particles. The parameter $g$ characterizes the strength of the non-linear coupling.

The dynamics associated with the energy functional (\ref{eq:GP_functional}) is described by the Lagrangian
\begin{equation}
 L[\psi]=\I \int \psi^*(\bs r,t)\;\partial_t\psi(\bs r,t) \;\D^D r -E[\psi].
\label{eq:action}
\end{equation}
The Euler-Lagrange equations then lead to the time-dependent Gross-Pitaevskii equation:
\begin{equation}
\I \partial_t\psi(\bs r,t)=-\frac{1}{2}\nabla^2\psi(\bs r,t)+g|\psi(\bs r,t)|^2 \psi(\bs r,t).
\label{eq:GPE_t}
\end{equation}


\subsection{Dimensional analysis for the soliton size}

Here we consider essentially the physics of an atomic Bose gas with the number of particles $N$ defined as
\begin{equation}
\int |\psi(\bs r,t)|^2\ {\rm d}^D r =N .
\label{eq:norm}
\end{equation}
The non-linear coefficient $g$ in (\ref{eq:GP_functional}) describes the strength of the interactions between the particles at the mean-field level. Here, we suppose that $g<0$, corresponding to an attractive interaction.

In order to obtain some intuition on the existence and stability of a soliton, we consider a stationary wave packet of size $\ell$, with a central density $N/\ell^D$, and  we perform a dimensional analysis of the energy per particle deduced from (\ref{eq:GP_functional}):
\begin{equation}
\frac{E(\ell)}{N}\ \sim\ \frac{1}{\ell^2}-\frac{N|g|}{\ell^D},
\label{eq:E_ell}
\end{equation} 
up to a numerical factor of order unity in front of each contribution.

\begin{figure}[t]
\begin{center}
\includegraphics{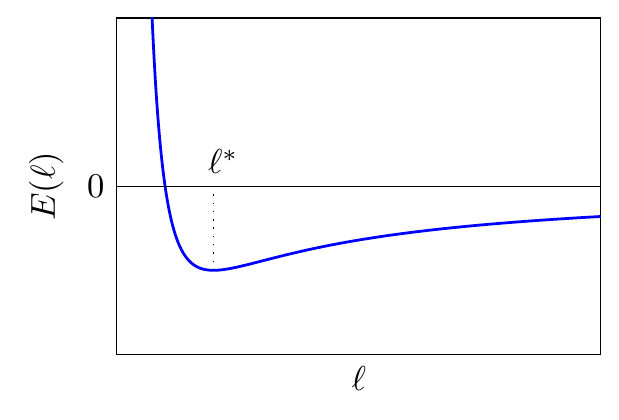}
\includegraphics{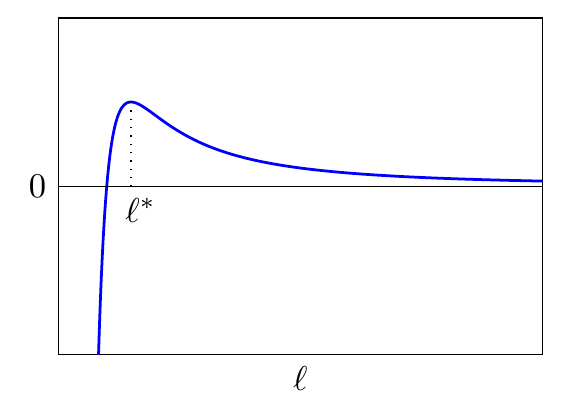}
\end{center}
\caption{Sketch of the energy per particle (\ref{eq:E_ell}) for a wave packet of size $\ell$ in 1D (left) and 3D (right).}
\label{fig:soliton_1_3}
\end{figure}

The variation of $E(\ell)$ deduced from (\ref{eq:E_ell}) is  shown in Fig.\,\ref{fig:soliton_1_3} for dimensions $D=1$ and $D=3$. The solution in the one-dimensional case is well known:  the attractive interaction term dominates at large $\ell$ and the kinetic energy dominates at small $\ell$. Therefore, there exists a size $\ell^*\sim 1/N|g|$ which corresponds to a stable equilibrium. In the context of cold atomic gases, such solitons were first observed in  \cite{Khaykovich:2002,Strecker:2002}. 

In 3D, the extremum of the energy functional, occurring for a length $\ell^* \sim N|g|$ is unstable. The expansion term due to the kinetic energy dominates for $\ell>\ell^*$, and the contraction term due to the interaction energy leads to a collapse for $\ell<\ell^*$. For small atom numbers, the fluid can be stabilized using an additional harmonic confinement, i.e. by adding a  component $\propto \ell^2$ to (\ref{eq:E_ell}) \cite{Bradley:1997a}. For large atom numbers, it leads to the celebrated phenomenon of "Bose Nova" \cite{donley2001dynamics}. 

By contrast to 1D and 3D, the two-dimensional case is critical, in the sense that no length scale emerges when we try to minimize  (\ref{eq:E_ell}). This is a consequence of the fact that the parameter $g$ is dimensionless in 2D, while it has the dimension of a length (resp. the inverse of a length) in 3D (resp. 1D). This absence of length scale associated with interactions in 2D is an illustration of the scale invariance of the action (\ref{eq:action}) in this case. 


\subsection{The 2D case: Townes profile}

For a more quantitative analysis of the extrema of the energy functional (\ref{eq:GP_functional}) in two dimensions, we turn to the time-independent Gross-Pitaevskii equation deduced from the minimization of (\ref{eq:GP_functional}):
\begin{equation}
-\frac{1}{2}\nabla^2\phi(\bs r)+g|\phi(\bs r)|^2 \phi(\bs r)=\mu\,\phi(\bs r)\qquad \qquad \int |\phi(\bs r)|^2\ {\rm d}^2 r =N.
\label{eq:GPE}
\end{equation}
Here the chemical potential $\mu$ is the Lagrange parameter  introduced to take into account the constraint on the normalization of $\phi$. Once $\phi(\bs r)$ is known, the function $\psi(\bs r,t)=\phi(\bs r)\,\E^{-\I \mu t}$ is a solution of the time-dependent Gross Pitaevskii equation (\ref{eq:GPE_t}).

It can been shown that (\ref{eq:GPE}) has physically acceptable solutions only for well-defined values of the product $Ng$. More precisely, these solutions have an isotropic density distribution and a phase dependence of the form $\exp( \I s \theta)$, $\theta$ being the polar angle and $s \in \mathbb{Z}$  the embedded vorticity \cite{sulem2007nonlinear}. For $s = 0$, the node-less solution of (\ref{eq:GPE}) exists only for $Ng=-5.85...$ \cite{1964_chiao,1964_talanov}, whereas solutions with one and two nodes require $Ng=-38.6...$ and $Ng=-97.9...$ respectively \cite{haus1966higher,yankauskas1966radial}.  The solution with no node (except in $r=0$) and with embedded vorticity $s = 1$ (resp. $s = 2$)  requires $Ng = -24.1...$ (resp. $Ng = -44.86...$) \cite{1992_Kruglov}.

\begin{figure}[t]
\begin{center}
\includegraphics{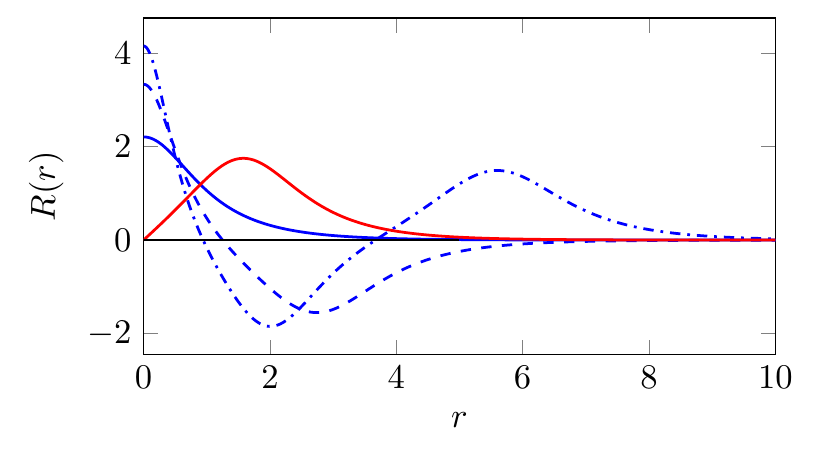}
\end{center}
\caption{Blue solid line: Radial profile of the Townes soliton, corresponding to the isotropic and nodeless solution of (\ref{eq:radial_equation}) in 2D. Dashed (resp. dash-dotted) line: Solution of (\ref{eq:radial_equation}) with one node (resp. two nodes). Red solid line: node-less solution with embedded vorticity $s = 1$. }
\label{fig:Townes_soliton}
\end{figure}

The $s = 0$ node-less solution corresponds to the celebrated Townes soliton. Its radial profile is given (after an arbitrary rescaling) by the positive solution $R(r)$ of the radial equation
\begin{equation}
 R'' +\frac{R'}{r} + R^3= R
\qquad\qquad
\int_0^{+\infty} R^2(r)\,2\pi r\;\D r=2\times 5.85...\, ,
\label{eq:radial_equation}
\end{equation}
and it is represented in figure \ref{fig:Townes_soliton}. Here, we have set $R(r)=\sqrt{2|g|}\,\phi(\bs r)$ in (\ref{eq:GPE}) and we have chosen the value $\mu=-1/2$ for the chemical potential (as we show below, this value can be chosen arbitrarily). 

This solitonic profile was found by Chiao, Garmire and Townes in \cite{1964_chiao} while they were searching for the transverse profile of a laser beam that could propagate without deformation in a non-linear medium, as a result of the competition between diffraction and self-focusing. This node-less profile does not exhibit any dynamical instability (i.e. no exponential growth for a small deformation from the ideal profile), by contrast to the solutions with one (or several) node(s) \cite{malkin1991elementary,1997_Firth_PhysRevLett.79.2450}, and we will thus focus on it in the following.  In the context of non-linear optics, the Townes soliton was observed by \cite{moll2003self} and we refer the reader to \cite{kartashov2019frontiers} for a thorough review of the numerous works performed on multi-dimensional self-trapping of laser beams. 

It is well known that the two-dimensional Gross-Pitaevskii action (\ref{eq:action}) is scale invariant and, more generally, conformal invariant. It presents a dynamical symmetry described by the SO(2,1) group  \cite{pitaevskii1997breathing}. This invariance leads to a series of remarkable properties for the weakly-interacting 2D Bose gas described by the classical-field approach (\ref{eq:GP_functional}) (for a review, see \cite{saint2019dynamical} and refs. in). In the context of solitons, the scale invariance corresponds to a special structure of the space of solutions of (\ref{eq:GPE}), as we show now. Let us assume that we choose $Ng=-5.85$ ; once we have determined a solution $\phi(\bs r)$ for a given chemical potential $\mu$, with the characteristic length $\mu^{-1/2}$ for this state, then we get a continuous family of solutions of the same equation (\ref{eq:GPE}),
\begin{equation}
\phi_\lambda(\bs r)=\lambda \phi(\lambda \bs r), \qquad \mu_\lambda=\lambda^2\mu,
\end{equation} 
containing the same atom number ($\int |\phi_\lambda|^2\,\D^2 r=\int |\phi|^2 \,\D^2 r=N$). Furthermore, all these solutions have zero energy for the functional (\ref{eq:GP_functional}), as shown in the next paragraph.


\subsection{Dynamics of 2D wave packets}
Among the many consequences of the dynamical symmetry associated to the Gross-Pitaevskii equation in 2D, we mention the following remarkable relation between the square of the spatial extension $\langle r^2\rangle(t)=\int r^2 |\psi(\bs r,t)|^2\,\D^2 r$ of the wave packet $\psi(\bs r,t)$, and its energy $E[\psi]$ which is a constant of motion \cite{pitaevskii1997breathing}: 
\begin{equation}
\frac{{\rm d}^2\langle r^2\rangle}{{\rm d}t^2}=\frac{4E}{m}.
\label{eq:variance_identity}
\end{equation}
Note that here we reintroduced explicitly the mass of the particles $m$ for clarity. From this result, we obtain immediately that a soliton, which has a time-independent size, must have a zero energy. More generally, we find the following implication:
\begin{equation}
E<0\quad \Rightarrow\quad \mbox{collapse.}
\label{eq:implication_1}
\end{equation}
Indeed, for a negative energy, we deduce from (\ref{eq:variance_identity}) that the variance $\langle r^2\rangle$ will vanish after a finite time (remember that $E$ is a constant of motion). The mathematical description of the collapsing dynamics has been widely studied and we refer the reader to \cite{sulem2007nonlinear} for more details. Note that the reciprocal statement of (\ref{eq:implication_1}) is not true: there exist collapsing wave packets that have a positive energy.

An important result regarding the possibility of collapse was obtained in \cite{weinstein1982nonlinear}. If the number of particles is small enough, one can be certain that there will be no collapse, i.e. no divergence of the density at any point in space. More precisely, one has the following result:
\begin{equation}
N|g| < \left(N|g|\right)_{\rm Townes}=5.85... \quad \Rightarrow\quad \mbox{no collapse.}
\label{eq:implication_2}
\end{equation} 
Here again, one should note that the reciprocal statement is not true: For arbitrary large values of $N|g|$, there exist wave packets which do not lead to collapse for $t>0$.

\begin{figure}[t]
\begin{center}
\includegraphics{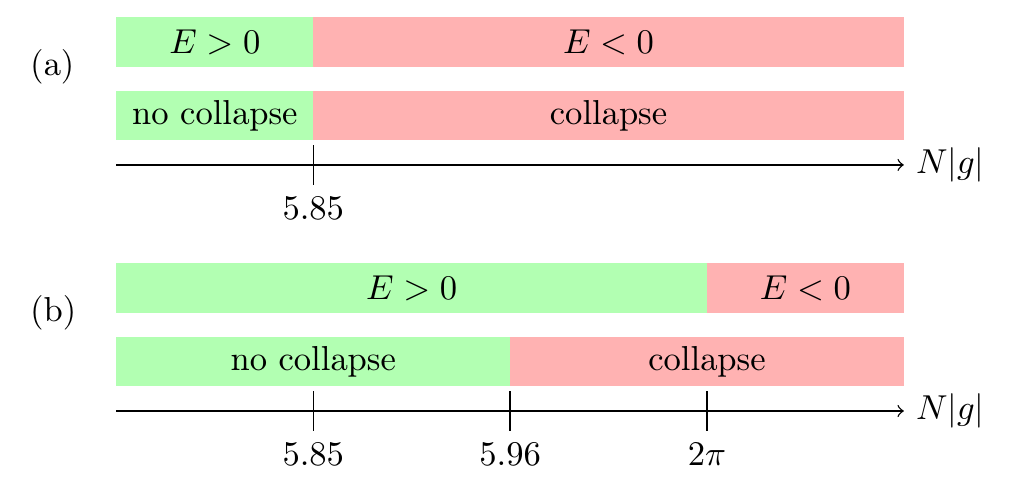}
\end{center}
\caption{The various scenarios in terms of expansion and collapse for (a) an initial profile proportional to the Townes profile $R(r)$;  (b) an initial Gaussian profile. Here,  $E>0$ implies an increase of $\langle r^2\rangle$ 
[see (\ref{eq:variance_identity}), with $\partial_t\langle r^2\rangle=0$ at $t=0$], and collapse means a divergence of the central density after a finite evolution time.}
\label{fig:Townes_vs_Gaussian}
\end{figure}

To show the subtlety of the 2D Gross-Pitaevskii equation, we illustrate these two implications by considering an wave packet that has initially either the Townes profile $R(r)$ or a gaussian shape (see also figure \ref{fig:Townes_vs_Gaussian}):\\

$-$ For an initial Townes profile, the situation is simple: if $N|g|<5.85$, the kinetic energy term is dominant in (\ref{eq:GP_functional}) so $E>0$. The size of the wave packet $(\langle r^2\rangle)^{1/2}$ will increase because of (\ref{eq:variance_identity}) and the central density will decrease because of (\ref{eq:implication_2}). If $N|g|=5.85$, the wave packet is stationary. If $N|g|>5.85$, the energy is negative and there will be a collapse after a finite evolution time.

\noindent $-$ For an initial Gaussian profile, one finds that the energy (\ref{eq:GP_functional}) is zero for $N|g|=2\pi$. One can thus infer that for  $N|g|>2\pi$, the energy is negative and it will lead to collapse, according to (\ref{eq:implication_1}). For $N|g|<5.85$, we know from (\ref{eq:implication_2}) that there will be no collapse. For intermediate values, $5.85<N|g|<2\pi$, the criteria given above do not allow one to conclude on the behavior of $\langle r^2\rangle$, nor on the possible occurence of a collapse, {\it i.e.}, a divergence of the central density. A numerical study has led to the following conclusions \cite{fibich2000critical}: The no-collapse region extends to $N|g|<5.96$ whereas a collapse occurs after a finite evolution time for $5.96<N|g|<2\pi$. This case is non-intuitive since, while the central density increases to infinity (by definition of the collapse), the size of the wave packet measured from $\langle r^2\rangle$ is also increasing since $E>0$. These two statements are simultaneously correct only because the wave packet gets strongly deformed with respect to its initial Gaussian shape.  


\subsection{Townes solitons with cold atoms}
\label{subsec:Townes_cold_atoms}

In 2021, two groups in Purdue and in Paris announced that they could prepare Townes solitons in a two-dimensional cold atomic gas  \cite{chen2021observation,2021_bakkali}. In both experiments, the two-dimensional character of the fluid was obtained thanks to a strong harmonic confinement of frequency $\omega_z$ along the third direction $z$. The atoms occupy essentially the ground state along this direction because both the residual temperature and the interaction energy per particle are small compared to $\hbar \omega_z$. The size of the single-particle ground state for the $z$ degree of freedom is $a_z=(\hbar/m\omega_z)^{1/2}$. In both experiments, $a_z$ was large compared to the 3D scattering length $a$, so that the 2D interaction parameter $g$ introduced above is simply $g=\sqrt{8\pi}a/a_z$ \cite{petrov2000bose,hadzibabic2011two}, which is a dimensionless quantity, as mentioned above.

The Purdue group \cite{chen2021observation} worked with  a a cesium gas and took advantage of a Feshbach resonance to suddenly switch the interaction strength $g$ from a positive to a negative value. This induces a modulational instability during which atoms group in several wave packets with various sizes and atoms numbers. The Purdue group could then show that these wave packets, when properly rescaled, all have a shape similar to $R(r)$, as expected for the Townes soliton. The atom number in each packet was in good agreement with the expected value $N=-5.85/g$. As this experiment did not involve an atomic mixture, its detailed description goes beyond the themes considered in this School and we will not discuss it further. 

\begin{figure}[t]
\begin{center}
\includestandalone[page=1]{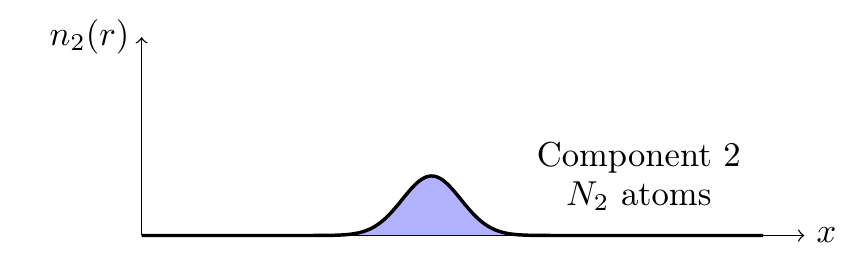}
\includestandalone[page=2]{two_components}
\end{center}
\caption{Strategy used in \cite{2021_bakkali} to generate an effective attractive interaction in the minority component 2. A wave packet of component 2 is immersed into a two-dimensional bath of component 1. Here, we show a cut of the density profiles along one direction ($x$) of the plane of atoms. The bath is assumed to be weakly depleted by the presence of the minority component. In this case, the interactions mediated by the bath add up to the "bare" interactions between atoms in component 2, so that the effective coupling parameter $g_{\rm eff}$, given in (\ref{eq:g_eff}), is negative (effective attractive interactions) although all $g_{ij}$ are positive.}
\label{fig:two_components}
\end{figure}

The experiment in Paris \cite{2021_bakkali} on the other hand was based on the specific properties of a mixture of Bose gases. More precisely, the soliton was formed in a deterministic way in a minority component, labelled hereafter 2, immersed in a bath of particles labelled 1. As we show below, when the bath density $n_1$ is everywhere much larger than the density of the minority component $n_2$,  the effective interaction strength for particles 2 can be written
\begin{equation}
g_{\rm eff}=g_{22}-\frac{g_{12}^2}{g_{11}}.
\label{eq:g_eff}
\end{equation} 
The experiment was performed with $^{87}$Rb atoms, and the components $(1,2)$ refer to the two hyperfine states $|F=(1,2), m_F=0\rangle$ (the so-called clock states). For these states, it is known that both intra and interspecies interaction strengths $g_{ij}$ are positive  \cite{van2002interisotope,altin2011optically}. However, for the choice of states given above, $g_{\rm eff}$ is slightly negative, which is the favourable situation to observe a Townes soliton for the minority component. In the experiment \cite{2021_bakkali}, the frequency of the confinement along $z$ is $\omega_z/2\pi=4.4\,$kHz, leading to $g_{\rm eff}\approx -0.0076$. The atom number corresponding to the Townes soliton is thus $N_{\rm T}\approx 770$.

\begin{figure}[t]
\begin{center}
\includegraphics[height=4cm]{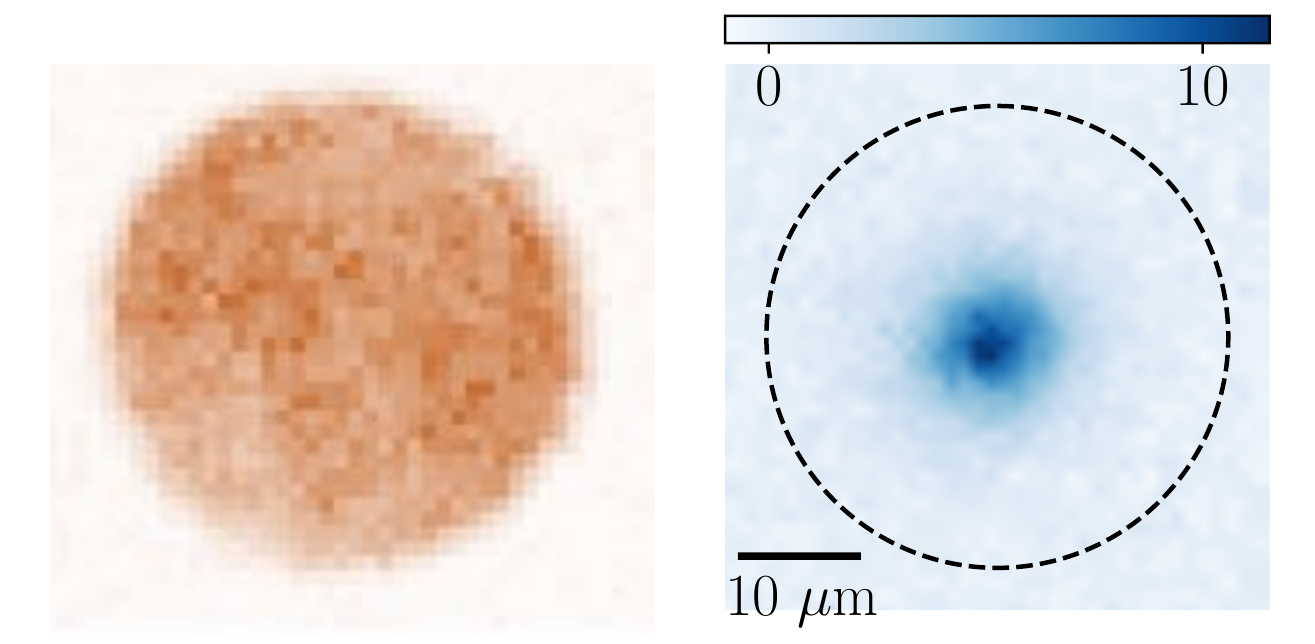}
\end{center}
\caption{Images of the two components used in \cite{2021_bakkali} to generate a Townes soliton. Left: Image of a uniform disk of $^{87}$Rb atoms in the hyperfine ground state $|1\rangle=|F=1,m=0\rangle$ (radius 20\,\si{\um}). The density is $\approx 100$ atoms/\si{\um}$^2$. Right: atomic cloud transferred to the hyperfine ground state $|2\rangle=|F=2,m=0\rangle$ by a stimulated Raman process. The color encodes the spatial density, which is $\approx 10$ atoms/\si{\um}$^2$ at the center of the cloud for this realization. The spatial profile of the laser beams used for the Raman process is optimized so that the density distribution of the atoms in $|2\rangle$ corresponds to the Townes profile $R^2(r/r_0)$, with $r_0$ of the order of a few \si{\um}. The atoms in $|1\rangle$ are still present in the right figure, but they are not resonant with the imaging laser beam and thus not detected. }
\label{fig:photos}
\end{figure}

The experiment described in \cite{2021_bakkali} starts with a uniform 2D gas of rubidium atoms prepared in state 1 inside a disk of radius $R_0 = 20$\,\si{\um} (figure \ref{fig:photos}). A pair of laser beams, at normal incidence with respect to the atomic plane, transfer a small fraction of atoms from state 1 to state 2 via a resonant stimulated Raman transition. The spatial profile of the laser beams is adjusted such that the density profile of the atoms transferred in $|2\rangle$ is very close to $R^2(r/r_0)$, where the characteristic length $r_0$ is on the order of a few \si{\um}, hence  $r_0\ll R_0$. The intensity of the laser beams and the exposure time are varied so as to explore the two regimes of atom numbers $N_2$ smaller and larger than the Townes threshold $N_{\rm T}$. It is kept low enough to ensure that the bath is weakly depleted (less than 20\%) at any point in space.

The experimental results confirmed the predictions for the condition of existence of the Townes soliton. When the atom number $N_2$ is lower than $N_{\rm T}$, the minority component expands and fill the whole disk. When the atom number approaches $N_{\rm T}$ (within $\pm 10\%$), the expansion stops and the wave packet formed by the atoms in $|2\rangle$ is essentially stationary on the time scale of the experiment (50 ms). For $N$ notably above $N_{\rm T}$, the wave packet contracts. This behavior is found for various values of the r.m.s. size of the initial minority wave packet (from 4 to 8 \si{\um}), which illustrates the scale invariance of the problem.


\section{The binary mixture approach to the Townes soliton}
\label{sec:binary_mixture}


\subsection{Coupled Gross-Pitaevskii equations}

We now turn to a quantitative description of the mediated interaction and the formation of the Townes soliton for the experimental protocol of \cite{2021_bakkali}, represented in figures \ref{fig:two_components}-\ref{fig:photos}. In the mean-field description corresponding to the Gross-Pitaevskii formalism, the steady-state of the binary system is obtained by solving the two equations:
\begin{eqnarray}
&& -\frac{1}{2} \nabla^2 \phi_1(\bs r) +\left[ g_{11} n_1(\bs r) +  g_{12} n_2(\bs r)  \right] \phi_1(\bs r) = \frac{m\mu_1}{\hbar^2} \phi_1(\bs r) \label{eq:twocomp_1}\\
&& -\frac{1}{2} \nabla^2 \phi_2(\bs r) +\left[ g_{12} n_1(\bs r) + g_{22} n_2 (\bs r)   \right] \phi_2(\bs r) =\frac{m\mu_2}{\hbar^2} \phi_2(\bs r) \label{eq:twocomp_2}
\end{eqnarray}
where $n_j(\bs r)=|\phi_j(\bs r)|^2$ represents the density of component $j$, and where we reintroduced explicitly $\hbar$ and the mass $m$ of each component (here assumed to be equal as in \cite{2021_bakkali}). 

We look for configurations such that the bath extends to infinity with the asymptotic density $n_{1,\infty}$. On the other hand, the minority component contains a finite number of atoms, $N_2=\int n_2(\bs r)\, \D^2r$ (see figure \ref{fig:two_components}). Far from the location of the minority component, we find $\phi_1\approx \sqrt{n_{1,\infty}}$ (up to a global phase factor), which provides the value $\mu_1=\frac{\hbar^2}{m}g_{11}n_{1,\infty}$. The  healing length of the bath is defined as 
\begin{equation}
\xibath\equiv\left(2g_{11}n_{1,\infty}\right)^{-1/2}
\label{eq:xi_1}
\end{equation}
and it gives the distance over which the bath "reacts" to the presence of a point defect. For the experiment of \cite{2021_bakkali}, $n_{1,\infty}\sim 100\,$\si{\um}$^{-2}$, $g_{11}=0.16$, leading to $\xibath\sim 0.2\,$\si{\um}. 


\subsection{The Thomas-Fermi approximation}
\label{subsec:TF_approx}

We suppose in this paragraph that the spatial scale over which $n_2(\bs r)$ varies is much larger than the healing length of the bath. We can then solve (\ref{eq:twocomp_1}) using the Thomas-Fermi approximation, which states that the bath density $n_1(\bs r)$ adjusts at any point to the minority density $n_2(\bs r)$, with a negligible cost in kinetic energy:
\begin{equation}
g_{11} n_1(\bs r) +  g_{12} n_2(\bs r) \approx \frac{m\mu_1}{\hbar^2}\qquad 
\Rightarrow\qquad n_1(\bs r) \approx n_{1,\infty}-\frac{g_{12}}{g_{11}}n_2(\bs r).
\label{eq:TF_approx}
\end{equation}
We then inject this result into the equation (\ref{eq:twocomp_2}) for $\phi_2$ and obtain:
\begin{equation}
 -\frac{1}{2} \nabla^2 \phi_2(\bs r) + g_{\rm eff} \,n_2(\bs r) \,\phi_2(\bs r) =\frac{m\mu_{\rm 2, eff}}{\hbar^2}\,\phi_2(\bs r)
\label{eq:GP_eff}
\end{equation}
where $g_{\rm eff}$ was defined in (\ref{eq:g_eff}) and where we set:
\begin{equation}
\mu_{\rm 2, eff}\equiv \mu_2-\frac{\hbar^2}{m}g_{12}n_{1,\infty}.
\label{eq:mu_2_eff}
\end{equation}
This is the result announced in the previous section: the effective interaction for the minority component is modified by the presence of the bath. The coupling constant $g_{\rm eff}$ can be negative, even when all $g_{ij}$'s are positive. Note that we require $\mu_{\rm 2, eff}< 0$ to ensure that $\phi_2$ tends to zero at infinity. 

We explained in \S\,\ref{sec:solitons_2D} that (\ref{eq:GP_eff}) can lead to the formation of a Townes soliton, provided the number of atoms $N_2$ is adjusted so that $|g_{\rm eff}|N_2=5.85$. Because of scale invariance, the soliton can have in principle an arbitrary size $\ell$. However, we remind that the use of the Thomas-Fermi approximation for the bath requires that $\ell \gg \xibath$. In addition, we should avoid to approach the regime of full depletion for the bath, so that the expression (\ref{eq:xi_1}) for the bath healing length $\xibath$ remains approximately correct even at the center of the minority wave packet. We discuss in \S\,\ref{subsec:beyond_TF} and in \S\,\ref{subsec:stron_depletion} how to handle situations (i) where $\xibath/\ell$ is not arbitrarily small and (ii) where the depletion of the bath can be quasi-total.


\subsection{Beyond Thomas-Fermi approximation: Breaking the scale invariance}
\label{subsec:beyond_TF}

In the Thomas-Fermi approach that we just described, one neglects any effect related to the kinetic energy of the bath, which is valid in the limit $\xibath/\ell\to 0$. To go one step further, we write the Gross-Pitaevskii equation (\ref{eq:twocomp_1}) for component 1 as
\begin{equation}
n_1(\bs r)=n_{1,\infty}-\frac{g_{12}}{g_{11}}n_2(\bs r)+\frac{\nabla^2 \sqrt{n_1(\bs r)}}{2 g_{11}\sqrt{n_1(\bs r)}},
\label{eq:exact_n1}
\end{equation}
where the last term of the right-hand side was neglected in the Thomas-Fermi approximation (\ref{eq:TF_approx}). 
We now assume that there are two small parameters in the problem:
\begin{equation}
\epsilon\equiv \frac{\xibath^2}{\ell^2}
,\qquad
\epsilon'\equiv\frac{n_{1,\infty}-n_{\rm 1,min}}{n_{1,\infty}}.
\label{eq:small_parameters}
\end{equation}
The hypothesis $\epsilon\ll 1$ was at the basis of the Thomas-Fermi approach, and the additional assumption $\epsilon'\ll 1$ corresponds to the case of a weakly depleted bath. In these conditions, we can approximate the last term in (\ref{eq:exact_n1}) by replacing $n_1(\bs r)$ by its Thomas-Fermi value, in the limit $\epsilon'\ll 1$:
\begin{equation}
\nabla^2 \sqrt {n_1(\bs r)} \approx \nabla^2\left[
\sqrt{n_{1,\infty}}-\frac{g_{12}}{2g_{11}}\frac{n_2(\bs r)}{\sqrt{n_{1,\infty}}}
\right].
\end{equation}
This leads to the following result: 
\begin{equation}
n_1(\bs r)\approx n_{1,\infty}-\frac{g_{12}}{g_{11}}n_2(\bs r)
-\frac{g_{12}}{4g_{11}^2n_{1,\infty}}\nabla^2 n_2(\bs r).
\label{eq:approached_n1}
\end{equation}
In this expression, the first correcting term $-\frac{g_{12}}{g_{11}}n_2(\bs r)$ with respect to the asymptotic value $n_{1,\infty}$ is of order $\epsilon'$ and the second correcting term is of order $\epsilon\epsilon'$. It is thus supposed to be a small correction compared to the first one, an hypothesis that should be checked for self-consistency at the end of the analysis.

Now we insert the expression (\ref{eq:approached_n1}) into the equation (\ref{eq:twocomp_2}) for $\phi_2$ and we obtain the single-component  equation:
\begin{equation}
 -\frac{1}{2} \nabla^2 \phi_2(\bs r) + g_{\rm eff} \,n_2(\bs r) \,\phi_2(\bs r) - \frac{\beta}{2} \left[ \nabla^2 n_2(\bs r) \right] \phi_2(\bs r) =\frac{m\mu_{\rm 2, eff}}{\hbar^2}\,\phi_2(\bs r)
\label{eq:Rosanov_eff}
\end{equation}
where $g_{\rm eff}$ and $\mu_{\rm 2, eff}$ were defined in (\ref{eq:g_eff}) and (\ref{eq:mu_2_eff}), respectively, and where we introduced
\begin{equation}
\beta = \frac{1}{ 2 n_{1,\infty}} \left(\frac{g_{12}}{g_{11}}\right)^2=|g_{\rm eff}|\,\xi_s^2 \qquad \mbox{with}\quad \xi_s=\frac{g_{12}}{g_{11}}\,\left(2 |g_{\rm eff}|n_{1,\infty}\right)^{-1/2}.
\end{equation}
Here $\xi_s$ is the so-called spin healing length, which is much larger than the bath healing length $\xibath$ when all coupling constants $g_{ij}$ are close to each other, hence $|g_{\rm eff}|\ll g_{ij}$. 

The modified Gross-Pitaevskii equation (\ref{eq:Rosanov_eff}) was introduced in \cite{2002_rosanov} to take into account a possible non-locality in the non-linear response of the medium in which the wave is propagating. Here, the non-local effects result from the fact that the density of the bath $n_1(\bs r)$ differs slightly from its Thomas-Fermi value, by a correction that involves a coarse-graining of $n_2$ on the spatial scale $\xibath$.   

To investigate the role of the additional term proportional to $\beta$ in (\ref{eq:Rosanov_eff}), we look at the energy per particle that is associated with each term of the left-hand side of this equation. As before, we consider a wave packet of extension $\ell$ that contains $N_2$ particles. The first two terms correspond to the kinetic energy and to the usual mean-field energy, respectively. Their contributions are both $\propto \ell^{-2}$ and they cancel each other at the Townes threshold $N_2=N_{\rm T}$ (i.e., when $g_{\rm eff}N_2=-5.85$). We therefore obtain an energy per particle that scales as (up to multiplicative constants of order unity):
\begin{equation}
\frac{E(\ell)}{N_2} \sim \frac{\hbar^2}{m}\left[\frac{g_{\rm eff}}{\ell^2}\left( N_{\rm T}-N_2\right)+\frac{\beta N_2}{2\ell^4}\right]
\label{eq:energy_Rosanov}
\end{equation}
that we should minimize with respect to $\ell$, for a given $N_2$.

For $N_2<\NT$, the two contributions in the bracket of (\ref{eq:energy_Rosanov}) are positive and the minimum is obtained for $\ell\to \infty$, which means that no stable wave packet exists in this case, as it was the case in the absence of the non-local correction. We recover the fact that for atom numbers smaller than the Townes threshold, the minority component tends to fill the whole accessible space, and thus overlaps with the bath even though the two components are non-miscible: this specific feature of 2D systems is due to the fact that the cost in kinetic energy would be too large to form finite-size domains of the minority component.

\begin{figure}[t]
\begin{center}
\includegraphics{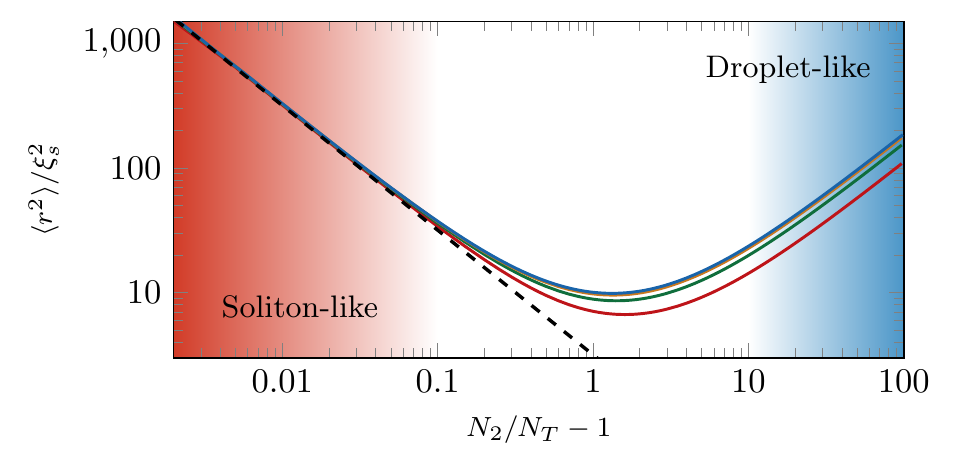}
\end{center}
\caption{Size of the minority component for various sets of interaction strengths $g_{11} = g_{22} < g_{12}$ close to the miscibility threshold. From top to bottom, $(g_{12} - g_{11}) / g_{11} = 1\%$ (blue solid line), $3\%$ (orange), $10\%$ (green),  $30\%$ (red). The black dashed line indicates the universal result \eqref{eq:size_Rosanov}, valid when $N_2 \rightarrow N_T^+$. In the strongly-depleted regime, the droplet has an approximately flat-top density profile $\approx n_{1,\infty}$, and thus a size $\langle r^2 \rangle$ that scales as $N_2$. In the limit $g_{12} \rightarrow g_{11}^+$, the curves converge towards the result expected from equation \eqref{eq:eq_sylvain}.}
\label{fig:size}
\end{figure}

For $N_2>\NT$, the minimum of $E(\ell)$ is obtained for a finite value of $\ell$ scaling as (see figure \ref{fig:size}):
\begin{equation}
\label{eq:size_Rosanov}
\ell \approx \alpha \,\xi_s \left(\frac{N_2}{N_2-N_{\rm T}}\right)^{1/2}
\end{equation}
where $\alpha$ is a numerical coefficient. A numerical resolution of (\ref{eq:Rosanov_eff}) confirms this scaling analysis and gives $\alpha\approx 1.8$ where $\ell$ designates the r.m.s.\,width of the wave packet. Therefore the addition in (\ref{eq:Rosanov_eff}) of the non-local term proportional to $\beta$ ensures that a stable solitary wave packet exists for any value of $N_2$ above the Townes threshold $N_{\rm T}$. This non-local term breaks scale invariance and brings the new length scale $\xi_s$ into the problem.  

\begin{figure}[t]
\begin{center}
\includegraphics{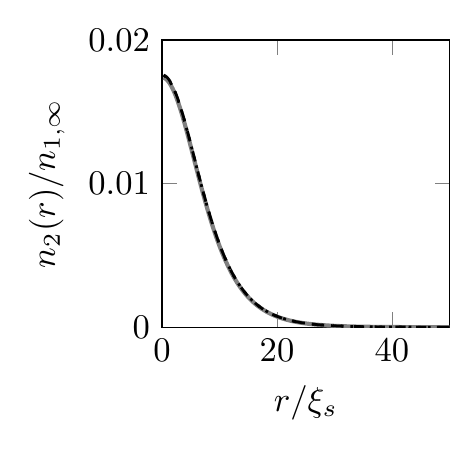}
\includegraphics{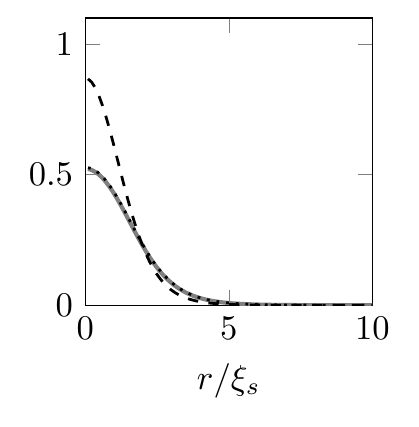}
\includegraphics{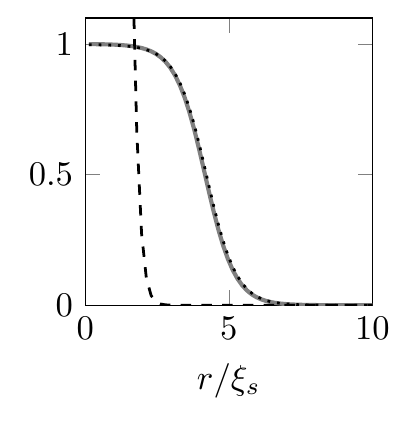}
\end{center}
\caption{Radial density profile for the minority component for $N_2=1.01\,N_{\rm T}$ (left), $1.5\,N_{\rm T}$ (middle) and $10\,N_{\rm T}$ (right). The calculations are made for $g_{22}= g_{11}$ and $g_{12}=1.01\,g_{11}$. The full line represents the solution of the two coupled Gross-Pitaevski equations (\ref{eq:twocomp_1}-\ref{eq:twocomp_2}). The dashed line and the dotted line are the solutions of the single component equations (\ref{eq:Rosanov_eff}) and (\ref{eq:eq_sylvain}), respectively. For this choice of the interaction parameters $g_{ij}$, the prediction derived from (\ref{eq:eq_sylvain}) nearly coincides with the solution of the two coupled Gross-Pitaevski equations for any value of $N_2$.
Figures adapted from \cite{bakkali2021testing}.}
\label{fig:three-profiles}
\end{figure}

To finish this analysis, we must check the validity of our expansion in terms of the small parameters $\epsilon$ and $\epsilon'$ introduced in (\ref{eq:small_parameters}).  The condition $\epsilon'\ll 1$ requires that $\frac{g_{12}}{g_{11}}n_2\ll n_{1,\infty}$. Using $n_2\sim N_2/\ell^2$, this condition is satisfied when $N_2-N_{\rm T}\ll \frac{g_{12}}{g_{11}}N_{\rm T}$. 
To simplify the discussion, consider the situation where all $g_{ij}$ are positive and close to each other, while $g_{\rm eff}$ is negative and notably smaller (by one order of magnitude) than $g$. The condition $\epsilon'\ll 1$ then imposes that $N_2$ should be chosen close to $\NT$. When this is the case, $\epsilon=\xibath^2/\ell^2$ is also much smaller than 1 and we expect the single-component equation (\ref{eq:Rosanov_eff}) to correctly describe the two-component mixture. This is confirmed by comparing the numerical solutions of (\ref{eq:twocomp_1}-\ref{eq:twocomp_2}) and of (\ref{eq:Rosanov_eff}), as shown in   figure \ref{fig:three-profiles}. 

\subsection{The strongly-depleted regime}
\label{subsec:stron_depletion}

The approach developed in the previous paragraph allowed us to describe the behavior of the two fluids by an effective single-component equation (see (\ref{eq:Rosanov_eff})). This approach is valid when the number of atoms in the minority component $N_2$ is just above the Townes threshold $N_{\rm T}=5.85/|g_{\rm eff}|$. When $N_2$ is notably above $N_{\rm T}$, this approach is not suited anymore and one has in principle to solve numerically the two coupled Gross-Pitaevskii equations (\ref{eq:twocomp_1}-\ref{eq:twocomp_2}). 

However, a single-component approach can still be developed in the case where all $g_{ij}$ are close to each other, as shown in the supplementary material of \cite{2021_bakkali} and the equation for the steady-state of component 2 reads:
\begin{equation}
\mu_2\, \phi_2(\boldsymbol r) =\frac{\hbar^2}{m}\left[ - \frac{1}{2} \nabla^2  + g_{\rm eff} n_2(\boldsymbol r)  + \frac{1}{2} \frac{ \nabla^2 \sqrt{ n_\infty - n_2(\boldsymbol r)} }{  \sqrt{ n_\infty - n_2(\boldsymbol r) } } \right]\phi_2(\boldsymbol r).
\label{eq:eq_sylvain}
\end{equation}
We will not present the derivation of this result here. We just signal that it is based on the fact that when all $g_{ij}$'s are close, the total density $n_1(\bs r)+n_2(\bs r)$ remains in all points close to $n_{1,\infty}$, which provides a small parameter (in addition to $\xi^2/\ell^2$) for a perturbation expansion.

We show in figure \ref{fig:three-profiles} that for $g_{22}=g_{11}$ and $g_{12}=1.01\,g_{11}$, the profiles obtained by solving the one-component equation (\ref{eq:eq_sylvain}) are in excellent agreement with the solution of the two coupled Gross-Pitaevskii equations for the whole range of values of $N/N_{\rm T}$. When $N\gg N_{\rm T}$ we reach a fully demixed regime, when the minority component forms a droplet inside the bath of component 1. The bath is nearly completely depleted at the position of this droplet, whose density is $\approx n_{1,\infty}$.

\subsection{Comparison with "quantum" droplets}

The wave packets that we consider in this lecture are stabilized under the action of pure mean-field interactions in a binary mixture. The physical situation is thus quite different from the quantum droplets considered by Petrov in \cite{2015_petrov} and implemented in
\cite{2017_tarruell,semeghini2018self,d2019observation,guo2021lee,Skov:2021_PhysRevLett.126.230404}\footnote{See also \cite{schmitt2016self,ferrier2016observation,chomaz2016quantum} for related experiments, performed in the presence of strong dipolar interactions. }, where  one takes advantage of beyond mean-field interactions to stabilize a 3D droplet \cite{bulgac2002dilute} that would otherwise collapse under the action of mean-field energy (for a review, see \cite{ferrier2019ultradilute}). 

The physical regimes considered here and in \cite{2015_petrov} are summarized in figure \ref{fig:axis_g}. In both cases, one assumes that each isolated component is stable ($g_{ii}>0$). The situation considered in \cite{2015_petrov} is close to the collapse point of the mixture, where the interspecies interaction is attractive ($g_{12}<0$) and compensates the intraspecies repulsion. By contrast, we consider here the case where $g_{12}$ is positive and close to the demixing threshold.   

For the 3D quantum droplets of size $\ell$ considered in  \cite{2015_petrov}, the three contributions to the energy per particle are (to be compared with (\ref{eq:energy_Rosanov}))
\begin{equation}
\mbox{kinetic: } \sim \frac{\hbar^2}{m\ell^2} \qquad \mbox{mean-field (M.F.): }\sim -|\delta G| n\qquad \mbox{beyond M.F.: }\sim\frac{m^{3/2}}{\hbar^3}\bar G^{5/2}n^{3/2}.
\label{eq:Petrov}
\end{equation}
In (\ref{eq:Petrov}), we assumed for simplicity that the 3D interaction parameters\footnote{In these notes, we write $G_{ij}$ with a capital letter for describing the 3D coupling constants, which as the dimension of energy$\times$volume, and we use the lower case letter $g_{ij}$ for the dimensionless 2D coupling constants.} $G_{11}= G_{22}$ and $N_1= N_2 \sim n\ell^3$, and we set $\bar G=(G_{11}G_{22})^{1/2}$, $\delta G= \frac{1}{2}(G_{12}+\sqrt{G_{11}G_{22}})<0$. In the regime where the kinetic energy can be neglected, the balance between the two other contributions leads to a state such that $n \bar a^3\sim (|\delta G|/\bar G)^2$, where $\bar a$ is the scattering length associated to $\bar G$. The passage from 3D to 2D or 1D, where the role of quantum fluctuations is enhanced and where the kinetic energy may play a decisive role, was studied in \cite{2016_petrov,cheiney2018bright,lavoine2021beyond,2021_sturmer}.

Let us finally mention the work of Naidon and Petrov \cite{naidon2021mixed}, in which the authors address the same situation as in this lecture, {\it i.e.}, $G_{12}$ positive and close to the demixing threshold $\sqrt{G_{11}G_{22}}$. By contrast to our study, they consider the case of notably different intraspecies interactions $G_{ii}$ and they show (for 1D, 2D and 3D) that beyond mean-field contributions can lead to the formation of bubbles with some population imbalance, immersed in a gas composed of only one of the two components. However, these effects seem out of reach for the experimental setup presented in \cite{2021_bakkali}. Indeed in that work, $g_{11}$ and $g_{22}$ were very close, and the parameter range over which the effect predicted in \cite{naidon2021mixed} is expected to occur is too narrow to be explored in practice.


\section{A microscopic view on the soliton formation}
\label{sec:microscopic}

So far, we analyzed the formation of a 2D soliton in the minority component using a mean-field analysis based on the two coupled Gross-Pitaevskii equations (\ref{eq:twocomp_1}-\ref{eq:twocomp_2}). In this section, we describe an alternative approach based on a polaron-type description of the interaction between the impurity atoms (component 2) and the bath (component 1). By contrast to \S\,\ref{sec:binary_mixture}, we will consider in this section the situation where the masses $m_1$ and $m_2$ may differ. We will indeed see that the analysis is notably simplified when $m_2\gg m_1$ (\S\,\ref{subsec:m2_gg_m1}), leading to a Yukawa-type interaction. The case $m_1=m_2$ will be then be addressed in \S\,\ref{subsec:Born_approx}.


\subsection{Reminder on the Bose polaron problem}
\label{subsec:Bose_polaron}

We consider a set of $N_2$ impurities of mass $m_2$ immersed in a  bath of bosons of mass $m_1$. We suppose that the bath is at zero temperature and that it can be described by the Bogoliubov approach. We denote $b_{\bs k},b_{\bs k}^\dagger$ the annihilation and creation operators for a Bogoliubov particle of momentum $\bs k$ and energy $\omega_k$. Since the gases studied in \cite{2021_bakkali} were in the quasi-2D regime, {\it i.e.}, the thickness of the gas was much larger than all 3D scattering lengths $a_{ij}$ \cite{petrov2000bose}, we perform here a 3D analysis of the impurity-bath interaction, which can subsequently provide the 2D coupling constant between the two components. We use periodic boundary conditions in a  volume $L^3$ so that the momenta $\bs k$ are quantized as $\bs k=\frac{2\pi}{L}(j_x,j_y,j_z)$ with   $j_\alpha \in \mathbb{Z}$. 

In the absence of coupling between the impurities and the bath, the hamiltonian reads
\begin{equation}
\hat H_0=\sum_{\bs k\neq 0}\hbar\omega_k\;b_{\bs k}^\dagger b_{\bs k} +\sum_{j=1}^{N_2}\frac{\hat{\bs p}_j^2}{2m_2}
\label{eq:hamiltonian_bath_impurities}
\end{equation} 
and the frequency of a Bogoliubov mode is
\begin{equation}
\hbar \omega_k= m_1c^2\,k\xi \,\left[2+(k\xi)^2\right]^{1/2} \qquad \mbox{with}\quad c=\frac{\hbar}{\sqrt 2\,m_1\xi}.
\label{eq:omegak}
\end{equation}

We use the Fr\"ohlich Hamiltonian\footnote{For an analysis that goes beyond the Fr\"ohlich Hamiltonian, see {\it e.g.} \cite{camacho2018bipolarons,camacho2018landau,van2018ground}.} to describe the interaction between the impurities and the bath (see for example 
\cite{grusdt2015new} 
for a derivation of this Hamiltonian):
\begin{equation}
\hat V=G_{12}n_1N_2+ \frac{1}{\sqrt {L^3}}\sum_{\bs k\neq 0} V_k \left(\sum_{j=1}^{N_2} \E^{\I \bs k\cdot \hat{\bs r}_j}\right)\left(b_{\bs k} +b_{-\bs k}^\dagger\right).
\label{eq:V_operator}
\end{equation}
Here $n_1$ stands for the 3D bath density in the absence of impurity (denoted $n_{1,\infty}$ in \S\,\ref{sec:binary_mixture}) and $G_{12}=2\pi\hbar^2 a_{12}/m_{12}$ is the 3D coupling constant between an impurity and an atom of the bath, with the reduced mass $m_{12}=m_1m_2/(m_1+m_2)$.
The first term in (\ref{eq:V_operator}) is the mean-field energy associated with the interaction between each impurity and the condensate. It appears at first order in $G_{12}$ and is a mere constant shift of all energies. The second term in (\ref{eq:V_operator}) describes the emission and absorption of a Bogoliubov quasi-particle due to the coupling with the impurities. The coefficient $V_k$ reads \cite{grusdt2015new}
\begin{equation}
V_k= G_{12}\sqrt{n_1} \left(\frac{(k\xi)^2}{(k\xi)^2+2} \right)^{1/4}.
\label{eq:Vk}
\end{equation}

We recall that in the expression (\ref{eq:V_operator}) of $\hat V$, the term $\bs k=0$ must be excluded from the sum, since we are interested in the role of Bogoliubov quasi-particles. The $\bs k=0$ term would correspond to a bath-impurity interaction that leaves the condensate unchanged, and such a term is already accounted for by the mean-field energy $G_{12}n_1$ for each impurity.


\subsection{Mediated interactions for $m_2\gg m_1$}
\label{subsec:m2_gg_m1}

We suppose in this paragraph that the mass  of an impurity $m_2$ is much larger than the mass of a bath particle $m_1$, so that the kinetic energy of the impurities in the Hamiltonian (\ref{eq:hamiltonian_bath_impurities}) can be neglected. We use second-order perturbation theory (with $G_{12}$ as the small parameter) to derive an effective interaction between the impurities.

We suppose that the impurities are prepared in a (properly symmetrized) state $|\Psi\rangle$ and that the bath is in its ground state, the Bogoliubov vacuum $|0_B\rangle$. The energy shift due to the coupling between the impurities and the bath is up to second order in $G_{12}$
\begin{equation}
\Delta E= G_{12}n_1N_2 - \sum'_\alpha \frac{|\langle \alpha|\hat V|\Psi,0_B\rangle|^2}{E_\alpha}
\label{eq:sum_Delta_E}
\end{equation}
where the sum runs over an orthonormal basis of eigenstates $|\alpha\rangle$ of $\hat H_0$, obtained by completing the initial state $|\Psi,0_B\rangle$. This initial state is of course excluded from the sum, as meant by the symbol $\displaystyle{\sum'_\alpha}$.

Using the expression (\ref{eq:V_operator}) for the impurity-bath coupling $\hat V$, it is clear that the states $|\alpha\rangle$ contributing to the sum (\ref{eq:sum_Delta_E}) contain one and only one Bogoliubov excitation $\bs k$. The energy $E_\alpha$ is $\hbar\omega_k$ since there is no energy coming from the impurities in $\hat H_0$, because of the assumption $m_2=\infty$. We denote $|\alpha\rangle=|\Psi',\bs k\rangle$ where $|\Psi'\rangle$ can be any of the states  chosen to form an orthogonal basis of the Hilbert space for the impurities. We then get
\begin{equation}
\Delta E= G_{12}n_1N_2 - \frac{1}{L^3} \sum_{\bs k\neq 0} \frac{V_k^2}{\hbar\omega_k} \;\sum_{|\Psi'\rangle} \left|
\langle \Psi'|\left(\sum_{j=1}^{N_2} \E^{\I \bs k\cdot \hat{\bs r}_j}\right)|\Psi\rangle
\right|^2
\end{equation}
The sum over $|\Psi'\rangle$ can be replaced by a closure relation and we are left with:
\begin{equation}
\Delta E= G_{12}n_1N_2 - \frac{1}{L^3} \sum_{\bs k\neq 0} \frac{V_k^2}{\hbar\omega_k}\;\langle \Psi|\left(N_2+\sum_{i\neq j}\E^{\I\bs k\cdot(\hat{\bs r}_j-\hat{\bs r}_i) }\right)|\Psi\rangle,
\label{eq:Delta_E_int}
\end{equation}
which can also be written
\begin{equation}
\Delta E= \left(G_{12} - \frac{1}{L^3} \sum_{\bs k\neq 0} \frac{V_k^2}{\hbar\omega_k}\right)n_1N_2 - \langle \Psi| \left[\sum_{i\neq j} \frac{1}{L^3} \sum_{\bs k\neq 0} \left(\frac{V_k^2}{\hbar\omega_k}\E^{\I\bs k\cdot(\hat{\bs r}_j-\hat{\bs r}_i) }\right)\right]|\Psi\rangle.
\label{eq:Delta_E_3}
\end{equation}

Two distinct effects appear on this result:
\begin{itemize}
 \item
A renormalization of the impurity-bath coupling constant:
\begin{equation}
G_{12} \quad \longrightarrow\quad G'_{12}=G_{12} - \frac{1}{L^3} \sum_{\bs k\neq 0} \frac{V_k^2}{\hbar\omega_k},
\label{eq:renormG}
\end{equation}
which corresponds to the inclusion of the second-order term in the Born expansion describing the interaction between a single impurity and the bath \cite{grusdt2015new}. We will not discuss it here, since we are rather interested in the interaction between two impurities mediated by the bath. 
 \item
An effective two-body interaction between the impurities corresponding to the potential
\begin{equation}
V_{\rm med}(\bs r)=-\frac{2}{L^3} \sum_{\bs k\neq 0} \left(\frac{V_k^2}{\hbar\omega_k}\E^{\I\bs k\cdot\bs r}\right).
\end{equation}
\end{itemize} 
With these two effects taken into account, (\ref{eq:Delta_E_3}) can be written
\begin{equation}
\Delta E= G'_{12}n_1N_2+\langle \Psi| \left(\frac{1}{2}\sum_{i\neq j}V_{\rm med}(\hat{\bs r}_i-\hat{\bs r}_j)\right)|\Psi\rangle.
\label{eq:Delta_E_with_G_prime}
\end{equation}

We now focus on the mediated potential $V_{\rm med}$. Using (\ref{eq:omegak}-\ref{eq:Vk}), we find
\begin{equation}
V_{\rm med}(\bs r)=-\frac{1}{L^3} \sum_{\bs k\neq 0} \frac{G_{12}^2}{G_{11}}\frac{\E^{\I\bs k\cdot\bs r}}{1+\frac{1}{2}(k\xi)^2} 
\label{eq:Vmed}
\end{equation}
where $G_{11}=4\pi \hbar^2 a_{11}/m_1$ is the bath interaction constant. The mediated potential can also be written in terms of its Fourier transform $\tilde V(\bs k)$:
\begin{eqnarray}
\bs k\neq 0 &:&\qquad  \tilde V_{\rm med}(\bs k)=-\frac{1}{L^3}\;\frac{G_{12}^2}{G_{11}}\frac{1}{1+\frac{1}{2}(k\xi)^2}
\label{Vmed_k_non_zero}\\ \bs k=0 &:&\qquad  \tilde V_{\rm med}(\bs k)=0,
\label{Vmed_k_zero}
\end{eqnarray}
where the reason for excluding the term $\bs k=0$ from the sum has been explained above. 

Note that (\ref{eq:Vmed}) can also be written
\begin{equation}
V_{\rm med}(\bs r)=\frac{1}{L^3}\frac{G_{12}^2}{G_{11}}-\frac{1}{L^3} \sum_{\bs k} \frac{G_{12}^2}{G_{11}}\frac{\E^{\I\bs k\cdot\bs r}}{1+\frac{1}{2}(k\xi)^2} 
=\frac{1}{L^3}\frac{G_{12}^2}{G_{11}}+V_{\rm Y}(r),
\label{eq:V_med_Yuk}
\end{equation}
where $V_{\rm Y}(r)$ is the attractive Yukawa potential \cite{2018_naidon}
\begin{equation}
V_{\rm Y}(r)=-\frac{1}{(2\pi)^3}\frac{G_{12}^2}{G_{11}}\int \frac{\E^{\I\bs k\cdot\bs r}}{1+\frac{1}{2}(k\xi)^2} \;\D^3 k=-\frac{G_{12}^2}{2\pi G_{11}\xi^2}\;\frac{\E^{-\sqrt 2\,r/\xi}}{r}.
\end{equation}
In two dimensions, this Yukawa potential is replaced by a potential involving the zeroth-order modified Bessel function of the first kind $K_0$, which varies as ${\rm e}^{-x}/\sqrt x$ at large distance \cite{bakkali2022cross}.


\subsection{Born approximation}
\label{subsec:Born_approx}

Let us suppose that the average distance between impurities is large compared to the healing length of the bath $\xi$. In the limit where the Yukawa potential is weak enough, we can describe it using Born approximation. More specifically, only $s$-wave collisions are relevant at low energy and we can replace $V_{\rm Y}$ by a (regularized) contact potential $\propto \delta(\bs r)$ with the scattering length $a_{\rm Y}$ defined as
\begin{equation}
\frac{2\pi \hbar^2 a_{\rm Y}}{m_{12}}=\int V_{\rm Y}(\bs r)\;\D^3 r=-\frac{G_{12}^2}{G_{11}}.
\label{eq:scatt_length_aY}
\end{equation}
This leads to the simple expression for the mediated potential (\ref{eq:V_med_Yuk}):
\begin{equation}
V_{\rm med}(\bs r)=\frac{G_{12}^2}{G_{11}}\left( \frac{1}{L^3} -\delta(\bs r)\right).
\label{eq:Med_contact}
\end{equation}

For low-energy scattering, a necessary condition for the validity of the Born approximation  is that $|a_{\rm Y}|$ should be much smaller than the range $\sim\xi$ of the Yukawa potential. Assuming that  $a_{12}\sim a_{11}$, one can check that this constraint is satisfied when the bath is in the weakly interacting regime, $\sqrt{na_{11}^3}\ll 1$, a condition that we assumed anyway in order to describe the bath within the Bogoliubov approach. 

In this section, we have assumed so far that the mass of the impurities $m_2$ is large compared to the mass of the bath particles $m_1$. When $m_2$ is comparable to $m_1$, the kinetic energy of the impurities has to be taken into account in the calculation of the energy shift $\Delta E$. To analyze the mediated two-body problem in this case, we consider a situation with two bosonic impurities. We assume that they are prepared in a (symmetrized) state of well defined momenta $(\bs p_a,\bs p_b)$. The energy shift of this state reads at second order in the coupling $G_{12}$  \cite{camacho2018landau}
\begin{equation}
E^{(2)}(\bs p_a,\bs p_b)=-\frac{V_k^2}{L^3}\left(
\frac{1}{ \hbar\omega_k +\Delta}+
\frac{1}{ \hbar\omega_k -\Delta}
\right) 
\label{eq:Delta_E_4}
\end{equation} 
where  $k=|\bs p_a-\bs p_b|/\hbar$ and $\Delta=({\bs p}_b^2-\bs p_a^2)/{2m_2}$. 
Note that we omitted in (\ref{eq:Delta_E_4}) the self-energy terms similar to (\ref{eq:renormG}) that renormalize the energy and the mass of each impurity, independently of the presence of the other one. Now, for a wave packet of extension $\ell\gg \xi$, the relevant momenta $\bs p_{a,b}$ are much smaller than $\hbar/\xi$ so that the dominant term in the denominators of (\ref{eq:Delta_E_4}) is the term linear in $k$, $\hbar \omega_k\approx \hbar c k$. We then recover the low-$k$ limit of the mediated potential (\ref{Vmed_k_non_zero}) that was obtained for $m_2\gg m_1$. In other words, the contact version of the mediated interaction (\ref{eq:scatt_length_aY}-\ref{eq:Med_contact}) can still be used when $m_1 \sim m_2$. The transposition of this result to the quasi-2D situation considered in \S\,\ref{sec:binary_mixture} then leads to the effective dimensionless coupling $g_{\rm eff}$ given in (\ref{eq:g_eff}). 

Note that this two-body problem can be equivalently addressed by calculating the scattering amplitude when two impurities with momenta $\bs p_a,\bs p_b$ collide in the presence of the bath \cite{camacho2018bipolarons}. The low-energy limit of this scattering amplitude provides the  scattering length associated with mediated interactions, which coincides with (\ref{eq:scatt_length_aY}).


\subsection{Bosonic vs. fermionic impurities}
\label{subsec:boson_vs_fermion}

In the previous paragraph, we evaluated the energy of a pair of bosonic impurities due to its interaction with the bath and got the negative result (\ref{eq:Delta_E_4}). It is straightforward to show that the same calculation performed for a pair of polarized fermionic impurities leads to an identical result in absolute value, but with the opposite sign, hence a positive energy shift. As explained in \cite{Yu:2012_PhysRevA.85.063616}, the result (\ref{eq:Delta_E_4}) for the induced interaction actually corresponds to an exchange term and thus takes opposite values for bosonic and fermionic impurities. 

Further insight in this problem is obtained by considering a cold assembly of $N_2$ impurities occupying the whole volume $L^3$, hence an impurity density $n_2=N_2/L^3$. For simplicity, we neglect any direct interactions between impurities ($a_{22}=0$). We use the contact form (\ref{eq:Med_contact}) of the mediated potential to evaluate the energy shift  (\ref{eq:Delta_E_with_G_prime}) and the effective binary interaction between impurities $\partial \mu_2/\partial n_2$  \cite{Yu:2012_PhysRevA.85.063616}:
\begin{itemize}
 \item
For polarized fermionic impurities, the contact term $\delta(\bs r)$ in (\ref{eq:Med_contact}) has no effect since the probability amplitude to have two fermions at the same location is null. More generally, if we come back to the spherically symmetric Yukawa potential, we know that it can only lead to scattering in odd partial wave channels for polarized fermions. At the very low temperature considered here, only $s$-wave channel is relevant and all odd channels have a negligible contribution. Therefore, we are left simply with:
\begin{equation}
\mbox{Fermions:}\qquad \Delta E=G'_{12}n_1N_2+\frac{N_2(N_2-1)}{2L^3}\frac{G_{12}^2}{G_{11}},
\end{equation}
leading to:
\begin{equation}
\mu_2=G'_{12}n_1+\frac{G_{12}^2}{G_{12}}n_2\qquad \Rightarrow\qquad 
\frac{\partial \mu_2}{\partial n_2}=\frac{G_{12}^2}{G_{11}}.
\label{eq:compressibility_fermions}
\end{equation}

 \item
For bosonic impurities occupying different momentum states $\bs p_1,\ldots,\bs p_{N_2}$ (hence a non-degenerate impurity gas), the contact interaction in (\ref{eq:Med_contact}) contributes both via a direct term and an exchange term, and one finds
\begin{equation}
\frac{\partial \mu_2}{\partial n_2}=\frac{G_{12}^2}{G_{11}}-2\frac{G_{12}^2}{G_{11}}=-\frac{G_{12}^2}{G_{11}},
\end{equation}
{\it i.e}, a result opposite to the fermionic one (\ref{eq:compressibility_fermions}).


\end{itemize}   


\section{Conclusions and outlook}
\label{sec:conclusions}

In this lecture, we studied the equilibrium configuration of a two-dimensional binary mixture of Bose gases, slightly beyond the demixing instability threshold. We first addressed the problem of a minority component immersed in a large bath of the other component from a mean-field point of view. Starting from coupled Gross-Pitaevskii equations, we could reduce this system to a single equation for the minority component. Under this framework, we explored topics ranging from the physics of Townes soliton and the breaking of scale invariance, to the very topical issue of droplet-like behaviors achieved with ultra-cold atoms. Next, we adopted a more microscopic approach and interpreted the minority component behavior as resulting from emergent interactions mediated by the bath elementary excitations, a problem related to the one of weakly-interacting Bose polarons. 

A natural extension of this work regards time-dependent situations. For instance, the study of near-equilibrium dynamics of the coupled fluids, with the search for possible eigenmodes of the systems as function of $N/N_{\rm T}$, shed new insights on the cross-over from Townes solitons to droplets \cite{bakkali2022cross}. Going one step further, one could study the collisional dynamics of such 2D wave packets, with possible behaviors ranging from near-elastic to strongly-inelastic (merging) interactions, thus enriching the now well-studied 1D situation (see e.g. \cite{2006_wieman, 2014_hulet, 2019_fattori, 2020_oberthaler}). It is our hope that the diversity of physical concepts discussed from this apparently simple settings will convince the reader of the wealth of phenomena encountered with such mixtures of quantum gases.


\acknowledgments

This work is supported by ERC TORYD, European Union's Horizon 2020 Programme (QuantERA NAQUAS project) and the ANR-18-CE30-0010 grant. B.B.H. acknowledges support from the Office of Naval Research. The authors thank Chlo\'e Maury, J\'er\^ome Beugnon and  Sylvain Nascimbene for their participation to the work that is described in these notes. J.D. also thanks Rudi Grimm and Cosetta Baroni for an insightful discussion during the summer school that led to the addition of \S\,\ref{subsec:boson_vs_fermion} to these notes.

\bibliographystyle{varenna}
\bibliography{Varenna}

\begin{thebibliography}{10}
\expandafter\ifx\csname url\endcsname\relax\def\url#1{\texttt{#1}}\fi
\expandafter\ifx\csname urlprefix\endcsname\relax\def\urlprefix{URL }\fi

\bibitem{2016_pitaevskii}
\NAME{Pitaevskii L. \atque Stringari S.}, \TITLE{Bose-{Einstein} {Condensation}
  and {Superfluidity}}, International {Series} of {Monographs} on {Physics}
  (Oxford University Press, Oxford) 2016.

\bibitem{2008_pethick}
\NAME{Pethick C.~J. \atque Smith H.}, \TITLE{Bose--Einstein Condensation in
  Dilute Gases}, 2nd Edition (Cambridge University Press) 2008.

\bibitem{2015_petrov}
\NAME{Petrov D.~S.}, \IN{{Physical Review Letters}}{115}{2015}{155302}.

\bibitem{1964_chiao}
\NAME{Chiao R.~Y., Garmire E. \atque Townes C.~H.}, \IN{{Physical Review
  Letters}}{13}{1964}{479}.

\bibitem{chen2021observation}
\NAME{Chen C.-A. \atque Hung C.-L.}, \IN{Physical Review
  Letters}{127}{2021}{023604}.

\bibitem{2021_bakkali}
\NAME{Bakkali-Hassani B., Maury C., Zou Y.-Q., Le~Cerf {\'E}., Saint-Jalm R.,
  Castilho P. C.~M., Nascimbene S., Dalibard J. \atque Beugnon J.},
  \IN{{Physical Review Letters}}{127}{2021}{023603}.

\bibitem{dauxois2006physics}
\NAME{Dauxois T. \atque Peyrard M.}, \TITLE{Physics of solitons} (Cambridge
  University Press) 2006.

\bibitem{Khaykovich:2002}
\NAME{Khaykovich L., Schreck F., Ferrari G., Bourdel T., Cubizolles J., Carr
  L.~D., Castin Y. \atque Salomon C.}, \IN{Science}{296}{2002}{1290}.

\bibitem{Strecker:2002}
\NAME{Strecker K.~E., Partridge G.~B., Truscott A.~G. \atque Hulet R.~G.},
  \IN{Nature}{417}{2002}{150}.

\bibitem{Bradley:1997a}
\NAME{Bradley C.~C., Sackett C.~A. \atque Hulet R.~G.}, \IN{{Physical Review
  Letters}}{78}{1997}{985}.

\bibitem{donley2001dynamics}
\NAME{Donley E.~A., Claussen N.~R., Cornish S.~L., Roberts J.~L., Cornell E.~A.
  \atque Wieman C.~E.}, \IN{Nature}{412}{2001}{295}.

\bibitem{sulem2007nonlinear}
\NAME{Sulem C. \atque Sulem P.-L.}, \TITLE{The nonlinear Schr{\"o}dinger
  equation: self-focusing and wave collapse}, Vol. 139 (Springer Science \&
  Business Media) 2007.

\bibitem{1964_talanov}
\NAME{Talanov V.~I.}, \IN{Izv. Vysshikh Uchebn. Zavedenii,
  Radiofiz.}{7}{1964}{}.

\bibitem{haus1966higher}
\NAME{Haus H.}, \IN{Applied Physics Letters}{8}{1966}{128}.

\bibitem{yankauskas1966radial}
\NAME{Yankauskas Z.}, \IN{Soviet Radiophysics}{9}{1966}{261}.

\bibitem{1992_Kruglov}
\NAME{Kruglov V.~I., Logvin Y.~A. \atque Volkov V.~M.}, \IN{Journal of Modern
  Optics}{39}{1992}{2277}.

\bibitem{malkin1991elementary}
\NAME{Malkin V. \atque Shapiro E.}, \IN{Physica D: Nonlinear
  Phenomena}{53}{1991}{25}.

\bibitem{1997_Firth_PhysRevLett.79.2450}
\NAME{Firth W.~J. \atque Skryabin D.~V.}, \IN{Phys. Rev.
  Lett.}{79}{1997}{2450}.

\bibitem{moll2003self}
\NAME{Moll K., Gaeta A.~L. \atque Fibich G.}, \IN{{Physical Review
  Letters}}{90}{2003}{203902}.

\bibitem{kartashov2019frontiers}
\NAME{Kartashov Y.~V., Astrakharchik G.~E., Malomed B.~A. \atque Torner L.},
  \IN{Nature Reviews Physics}{1}{2019}{185}.

\bibitem{pitaevskii1997breathing}
\NAME{Pitaevskii L. \atque Rosch A.}, \IN{Physical Review A}{55}{1997}{R853}.

\bibitem{saint2019dynamical}
\NAME{Saint-Jalm R., Castilho P.~C., Le~Cerf {\'E}., Bakkali-Hassani B., Ville
  J.-L., Nascimbene S., Beugnon J. \atque Dalibard J.}, \IN{Physical Review
  X}{9}{2019}{021035}.

\bibitem{weinstein1982nonlinear}
\NAME{Weinstein M.~I.}, \IN{Communications in Mathematical
  Physics}{87}{1982}{567}.

\bibitem{fibich2000critical}
\NAME{Fibich G. \atque Gaeta A.~L.}, \IN{Optics letters}{25}{2000}{335}.

\bibitem{petrov2000bose}
\NAME{Petrov D., Holzmann M. \atque Shlyapnikov G.}, \IN{Physical Review
  Letters}{84}{2000}{2551}.

\bibitem{hadzibabic2011two}
\NAME{Hadzibabic Z. \atque Dalibard J.}, \IN{La Rivista del Nuovo
  Cimento}{34}{2011}{389}.

\bibitem{van2002interisotope}
\NAME{Van~Kempen E., Kokkelmans S., Heinzen D. \atque Verhaar B.},
  \IN{{Physical Review Letters}}{88}{2002}{093201}.

\bibitem{altin2011optically}
\NAME{Altin P., McDonald G., Doering D., Debs J., Barter T., Close J., Robins
  N., Haine S., Hanna T. \atque Anderson R.}, \IN{New Journal of
  Physics}{13}{2011}{065020}.

\bibitem{2002_rosanov}
\NAME{Rosanov N.~N., Vladimirov A.~G., Skryabin D.~V. \atque Firth W.~J.},
  \IN{{Physics Letters A}}{293}{2002}{45}.

\bibitem{bakkali2021testing}
\NAME{Bakkali-Hassani B.}, \TITLE{Testing scale invariance in a two-dimensional
  {B}ose gas: preparation and characterization of solitary waves}, Ph.D.
  thesis, Sorbonne universit{\'e} (2021).

\bibitem{2017_tarruell}
\NAME{Cabrera C.~R., Tanzi L., Sanz J., Naylor B., Thomas P., Cheiney P. \atque
  Tarruell L.}, \IN{Science}{359}{2018}{301}.

\bibitem{semeghini2018self}
\NAME{Semeghini G., Ferioli G., Masi L., Mazzinghi C., Wolswijk L., Minardi F.,
  Modugno M., Modugno G., Inguscio M. \atque Fattori M.}, \IN{Physical review
  letters}{120}{2018}{235301}.

\bibitem{d2019observation}
\NAME{D'Errico C., Burchianti A., Prevedelli M., Salasnich L., Ancilotto F.,
  Modugno M., Minardi F. \atque Fort C.}, \IN{Physical Review
  Research}{1}{2019}{033155}.

\bibitem{guo2021lee}
\NAME{Guo Z., Jia F., Li L., Ma Y., Hutson J.~M., Cui X. \atque Wang D.},
  \IN{Physical Review Research}{3}{2021}{033247}.

\bibitem{Skov:2021_PhysRevLett.126.230404}
\NAME{Skov T.~G., Skou M.~G., J\o{}rgensen N.~B. \atque Arlt J.~J.},
  \IN{{Physical Review Letters}}{126}{2021}{230404}.

\bibitem{schmitt2016self}
\NAME{Schmitt M., Wenzel M., B{\"o}ttcher F., Ferrier-Barbut I. \atque Pfau
  T.}, \IN{Nature}{539}{2016}{259}.

\bibitem{ferrier2016observation}
\NAME{Ferrier-Barbut I., Kadau H., Schmitt M., Wenzel M. \atque Pfau T.},
  \IN{{Physical Review Letters}}{116}{2016}{215301}.

\bibitem{chomaz2016quantum}
\NAME{Chomaz L., Baier S., Petter D., Mark M., W{\"a}chtler F., Santos L.
  \atque Ferlaino F.}, \IN{Physical Review X}{6}{2016}{041039}.

\bibitem{bulgac2002dilute}
\NAME{Bulgac A.}, \IN{{Physical Review Letters}}{89}{2002}{050402}.

\bibitem{ferrier2019ultradilute}
\NAME{Ferrier-Barbut I.}, \IN{Physics Today}{72}{2019}{46}.

\bibitem{2016_petrov}
\NAME{Petrov D.~S. \atque Astrakharchik G.~E.}, \IN{{Physical Review
  Letters}}{117}{2016}{100401}.

\bibitem{cheiney2018bright}
\NAME{Cheiney P., Cabrera C., Sanz J., Naylor B., Tanzi L. \atque Tarruell L.},
  \IN{{Physical Review Letters}}{120}{2018}{135301}.

\bibitem{lavoine2021beyond}
\NAME{Lavoine L. \atque Bourdel T.}, \IN{Physical Review A}{103}{2021}{033312}.

\bibitem{2021_sturmer}
\NAME{St\"urmer P., Tengstrand M.~N., Sachdeva R. \atque Reimann S.~M.},
  \IN{Physical Review A}{103}{2021}{053302}.

\bibitem{naidon2021mixed}
\NAME{Naidon P. \atque Petrov D.}, \IN{Physical Review
  Letters}{126}{2021}{115301}.

\bibitem{camacho2018bipolarons}
\NAME{Camacho-Guardian A., Ardila L.~P., Pohl T. \atque Bruun G.~M.},
  \IN{{Physical Review Letters}}{121}{2018}{013401}.

\bibitem{camacho2018landau}
\NAME{Camacho-Guardian A. \atque Bruun G.~M.}, \IN{Physical Review
  X}{8}{2018}{031042}.

\bibitem{van2018ground}
\NAME{Van~Loon S., Casteels W. \atque Tempere J.}, \IN{Physical Review
  A}{98}{2018}{063631}.

\bibitem{grusdt2015new}
\NAME{Grusdt F. \atque Demler E.}, \TITLE{Quantum Matter at Ultralow
  Temperatures}, Vol. 191 of \emph{Proceedings of the International School of
  Physics "Enrico Fermi} (IOS Press Amsterdam) 2015, Ch. New theoretical
  approaches to Bose polarons, p. 325.

\bibitem{2018_naidon}
\NAME{Naidon P.}, \IN{{Journal of the Physical Society of
  Japan}}{87}{2018}{043002}.

\bibitem{bakkali2022cross}
\NAME{Bakkali-Hassani B., Maury C., Stringari S., Nascimbene S., Dalibard J.
  \atque Beugnon J.}, \IN{arXiv preprint arXiv:2207.06939}{}{2022}{}.

\bibitem{Yu:2012_PhysRevA.85.063616}
\NAME{Yu Z. \atque Pethick C.~J.}, \IN{Phys. Rev. A}{85}{2012}{063616}.

\bibitem{2006_wieman}
\NAME{Cornish S.~L., Thompson S.~T. \atque Wieman C.~E.}, \IN{{Physical Review
  Letters}}{96}{2006}{170401}.

\bibitem{2014_hulet}
\NAME{Nguyen J. H.~V., Dyke P., Luo D., Malomed B.~A. \atque Hulet R.~G.},
  \IN{{Nature Physics}}{10}{2014}{918}.

\bibitem{2019_fattori}
\NAME{Ferioli G., Semeghini G., Masi L., Giusti G., Modugno G., Inguscio M.,
  Gallem\'{\i} A., Recati A. \atque Fattori M.}, \IN{{Physical Review
  Letters}}{122}{2019}{090401}.

\bibitem{2020_oberthaler}
\NAME{Lannig S., Schmied C.-M., Pr\"ufer M., Kunkel P., Strohmaier R., Strobel
  H., Gasenzer T., Kevrekidis P.~G. \atque Oberthaler M.~K.}, \IN{{Physical
  Review Letters}}{125}{2020}{170401}.

\end{thebibliography}

\end{document}